# Sensitivity Analysis of the MCRF Model to Different Transiogram Joint Modeling Methods for Simulating Categorical Spatial Variables

Bo Zhang, Weidong Li, Chuanrong Zhang

*Department of Geography and Center for Environmental Sciences and Engineering, University of Connecticut, Storrs, CT 06269-4148, USA*

Bo Zhang: bozhang@uconn.edu
Weidong Li: weidong.li@uconn.edu
Chuanrong Zhang: chuanrong.zhang@uconn.edu

**Abstract**

Markov chain geostatistics is a methodology for simulating categorical fields. Its fundamental model for conditional simulation is the Markov chain random field (MCRF) model, and its basic spatial correlation measure is the transiogram. There are different ways to get transiogram models (i.e., continuous-lag transiograms) for MCRF simulation based on sample data and expert knowledge: linear interpolation method, mathematical model joint-fitting method, and a mixed method of the former two. Two case studies were conducted to show how simulated results, including optimal prediction maps and simulated realization maps, would respond to different sets of transiogram models generated by the three different transiogram jointing modeling methods. Results show that the three transiogram joint modeling methods are applicable; the MCRF model is generally not very sensitive to the transiogram models produced by different methods, especially when sample data are sufficient to generate reliable experimental transiograms; and the differences between overall simulation accuracies based on different sets of transiogram models are not significant. However, some minor classes show obvious improvement in simulation accuracy when theoretical transiogram models (generated by mathematical model fitting with expert knowledge) are used for minor classes. In general, this study indicates that the methods for deriving transiogram models from experimental transiograms can perform well in conditional simulations of categorical soil variables when meaningful experimental transiograms can be estimated. Employing mathematical models for transiogram modeling of minor classes provides a way to incorporate expert knowledge and improve the simulation accuracy of minor classes.

# 1. Introduction

Categorical spatial variables, such as soil type, land cover class, hydrofacies, and lithofacies, usually show complex spatial patterns. Assessing the spatial distributions of spatial categories in large areas with field survey is often difficult; however, it is critical to understanding their spatial patterns for environmental studies and land management. For soil types, traditionally this information could be obtained from experts' judgment or non-statistical methods such as Voronoi polygons [1]. These methods did not provide quantitative measurements of the spatial patterns nor the probability of the predictions. Geostatistical tools may provide a way of quantitatively describing the spatial or spatiotemporal patterns of soil categorical spatial variables.

In the past few decades, geostatistical methods have been widely employed to characterize the spatial patterns of soil properties [8, 25]. For dealing with soil categorical variables, such as soil types, soil textural classes, and soil water table classes, the geostatistical methods that may be considered include both indicator kriging methods and non-kriging methods. For example, the indicator kriging approach was introduced for interpolating soil water table classes [2], the Markov chain random field (MCRF) approach was introduced for simulating soil types and soil textural classes [17, 28], and the multiple-point geostatistics was used for simulating soil material classes [22]. MCRF was initially proposed for solving the small class underestimation problem of a coupled Markov chain model and extending 1-D Markov chain model to a generalized geostatistical approach, as described in [11] and [14]. A generalized formula of the MCRF idea for local conditional probability distribution estimation was derived using the idea of sequential Bayesian updating on nearest data within a neighborhood. The spatial conditional independence assumption for nearest neighbors was also proposed by Li [11] and used to simplify the full formula of the MCRF model that contains multi-point likelihood terms into a simplified formula that contains only transition probability terms, so that the MCRF model can be implemented with transiogram models. Therefore, the MCRF model is a nonlinear spatial Bayesian updating model for simulating complex categorical fields. Sequential simulation algorithms have been developed based on the MCRF model and the coMCRF model [13, 17] for simulating categorical soil variables and post-processing land cover classification [18, 30]. Compared with other geostatistical methods for simulating categorical spatial variables, such as sequential indicator simulation, MCRF simulation algorithms can generate more converged (usually polygon-shaped) and more rational categorical patterns with higher accuracy [13].

The transiogram concept was suggested by Li [12] as a unique name for expressing transition probability-lag diagrams, based on related pioneer studies [3, 7, 20, 24] and the variogram concept used in conventional geostatistics (i.e., kriging methods) [21]. The transiogram was proposed as a spatial correlation measure for MCRF modeling because it can be practically and conveniently estimated from sample data of categorical spatial variables. A transiogram refers to a transition probability curve over increasing distance lag [15]. To a large extent, transiograms are similar to indicator variograms [3, 6]; but compared with indicator variograms, transiograms can measure the cross-correlations of spatial categories with more features. There are some differences between them: Transiograms are intuitively more interpretable, and cross-transiograms are asymmetrical and can be unidirectional. In addition, transiograms in an idealized form (i.e., stationary Markovian) can be directly computed from a transition probability matrix if available [12]. More recently, Zhai et al. [27] demonstrated that transiograms can be a valuable graphic metric for characterizing the spatial patterns of landscape classes and the peak height ratio of cross-transiograms may be a good index for expressing the neighboring strength of class pairs.

To perform simulations using the MCRF approach, transiogram models (i.e., continuous-lag transition probably diagrams or fitted mathematical models) are needed. Similar to experimental variograms, experimental transiograms, which are directly estimated from sample data, are not lag-continuous. Thus, a transiogram modeling process is necessary to get continuous-lag transiogram models that can provide required transition probability values at any lag values. To meet the constraint conditions for transition probabilities such as the nonnegative and summing-to-unity, joint modeling is required. Two transiogram joint modeling methods were suggested for obtaining a whole set of transiogram models by fitting with experimental transiograms to implement the MCRF model - one is the linear interpolation method and the other is the mathematical model joint fitting method, both of which have been tested to be reliable under most conditions [15]. These two methods both have their merits and disadvantages. While the linear interpolation method is very efficient, it may not be a suitable choice for very small and sparse sample datasets and for minor classes with only a few sample points, because experimental transiograms under these situations may not be reliable and users cannot adjust the empirical models from linear interpolation. The mathematical model joint fitting method is relatively time-consuming, especially when the number of transiograms is large, but it is feasible to generate a set of reliable transiogram models for very small sparse sample datasets or minor classes because users can adjust the theoretical models through setting parameters (e.g., sill, range, and model type) based on expert knowledge and the trends of the probably-unreliable experimental transiograms. The third transiogram joint modeling method may consider combining the two methods mentioned above together, so that their merits may be kept and their disadvantages may be avoided.

It should be mentioned that idealized transiograms (directly calculated from a transition probability matrix with the implied Markovian assumption of spatial data) also can be used as transiogram models if available, because linear interpolation may be applied to make them lag-continuous if they are not. However, due to their over-idealized nature (because real data are rarely Markovian), it may not be a good choice for simulating complex patterns in horizontal dimensions. Studies [16, 27] have clearly shown that large discrepancies may exist between some experimental cross-transiograms and corresponding idealized cross-transiograms of landscape classes, even at the low lag section. In addition, the continuous-lag Markov chain models proposed by Carle and Fogg [4], which was grouped into idealized transiograms by Li [12] because they are also based on stationary Markovian assumption, may be more capable in subsurface characterization if accurate transition probabilities at multiple lag values can be estimated from borehole data and incorporated into the models.

The objective of this study is to examine the sensitivity of the MCRF model to the three transiogram joint modeling methods as well as the pros and cons of the three methods in simulating the spatial distribution of surface soil spatial variables – soil types. The mixed transiogram joint modeling method that combines the linear interpolation method and the mathematical model fitting method is suggested. The MCRF sequential simulation (MCSS) algorithm with a quadrantal neighborhood and spiral search is used for conditional simulation, as this neighborhood choice was proven to be the optimal one in normal situations [19]. Two case studies are conducted to test how the optimal prediction maps and realizations maps would differ with different transiogram model sets obtained using different joint modeling methods.

## 2. Methods

The MCRF theory was based on three basic ideas – spatial sequential Bayesian updating, spatial conditional independence assumption, and the transiogram spatial metric. It extended a single Markov chain into a multi-D geostatistical model for categorical data by conditioning a single (continuous-path or jumping) Markov chain on local nearest data. The MCRF model determines the local conditional probability distribution of a random variable at an unobserved location through Bayesian updating at the individual nearest data level in different directions in a neighborhood. The spatial sequential Bayesian updating essentially turns the locally-conditioned Markov chain into a spatial Bayesian network over spatial data by following Bayes' theorem, which was found to be consistent with the nonspatial Bayesian networks proposed in artificial intelligence [10, 23], as proven with details in [19]. After the spatial conditional independence assumption is applied, the MCRF model is simplified into a probability expression that consists of only transition probabilities over spatial lags [11, 19]. Thus, the basic spatial correlation measure used in the MCRF model is the transiogram (i.e., transition probability-lag diagram). The MCRF model has nothing to do with maximum entropy, and it is also not a conventional Markov random field model based on cliques and Gibbs distribution.

### 2.1 Transiogram and joint modeling methods

*2.1.1 Transiogram concept*

Based on Markov chain theory, relevant pioneer studies [3, 20, 24] and the variogram concept, the transiogram is defined as a transition probability-lag function theoretically and a transition probability-lag curve visually [12]:

$$p_{ij}(\mathbf{h}) = \Pr(Z(\mathbf{u} + \mathbf{h}) = j | Z(\mathbf{u}) = i)$$

where $p_{ij}(\mathbf{h})$ is the transition probability from state (i.e., class for a categorical spatial variable) $i$ to state $j$ with a lag of $\mathbf{h}$, and $i$ and $j$ are called the tail class and the head class, respectively. With increasing lag value, the $p_{ij}(\mathbf{h})$ forms a continuous diagram. $p_{ii}(\mathbf{h})$ represents an auto-transiogram of a single class and $p_{ij}(\mathbf{h})$ ($i \neq j$) represents a cross-transiogram of a class pair. Given the second-order stationarity assumption, the $p_{ij}(\mathbf{h})$ only depends on the lag $\mathbf{h}$ but not the location $\mathbf{u}$ [12]. Transiograms can be omnidirectional, bidirectional or unidirectional, depending on how it estimated. The transiogram concept here is not attached to indicator variogram and joint probability, so that it does not lose the legitimacy of Markov chain transition probability that it may be unidirectional and may be derived from a transition probability matrix.

Similar to variogram, transiograms can be directly estimated from sample data and such kind of transiograms is called "experimental (or empirical) transiograms". Transiograms directly estimated from an exhaustive map dataset may be regarded as a special type of experimental transiograms. Interestingly, transiograms even can be calculated from a transition probability matrix based on the Markov chain theory if it is available; and this kind of transiograms are called "idealized transiograms", due to the facts that they are very smoothing curves and imply the stationary first-order Markovian assumption on the data that produce the transition probability matrix. As demonstrated by Zhai et al. [27], transiograms can be used as a graphic metric to provide valuable spatial pattern information of landscapes that cannot be captured by traditional landscape metrics. Compared with indicator variogram or covariance, transiograms can express more features of spatial correlations, as the cross correlation of a pair of classes is represented by two cross-transiograms.

When calculating experimental transiograms from a sample dataset, the computer program counts the number of transitions from one class to either itself or another class at each lag distance specified by users, and divides the count by the sum of the numbers of transitions from the class to all classes at the lag distance as follows:

$$\hat{p}_{ij}(\mathbf{h}) = \frac{F_{ij}(\mathbf{h})}{\sum_{k=1}^{n} F_{ik}(\mathbf{h})}$$

where $F_{ij}(\mathbf{h})$ is the count of transitions from state $i$ to $j$ over the lag $\mathbf{h}$. Here $\mathbf{h}$ can be bidirectional or unidirectional but is often omni-directional for simulating landscape categories in the horizontal dimensions. However, to get reliable transition probability values from a limited number of sample data, $\mathbf{h}$ is usually replaced by a value interval (i.e., ($\mathbf{h}$-0.5$\Delta\mathbf{h}$, $\mathbf{h}$+0.5$\Delta\mathbf{h}$]) set by users.

For each categorical dataset with *n* classes, there are *n²* transiograms to describe the auto and cross correlations (or intra and inter class dependences) of the *n* classes, which forms a *n×n* transiogram matrix (similar to a transition probability matrix), due to the asymmetric property of transition probabilities (i.e., $p_{ij}(\mathbf{h}) \neq p_{ji}(\mathbf{h})$). However, if transition probabilities are estimated omni-directionally or bi-directionally, we have $p_{ij}(\mathbf{h})p_i = p_{ji}(\mathbf{h})p_j$, which means reversibility. Transition probabilities (or transiograms) have the constraint conditions of being nonnegative and sum-to-unity in each matrix row at any lag value [12].

*2.1.2 Linear interpolation joint modeling*

An experimental transiogram is normally a discrete sequence of transition probabilities at different lag values. To obtain a transiogram model with continuous lag that can provide transition probability values at any lag value, a fast and straightforward method is to linearly interpolate the experimental transiogram into a continuous curve using the following equation:

$$p_{ij}(\mathbf{h}) = \frac{\hat{p}_{ij}(\mathbf{h}_k)(\mathbf{h}_{k+1} - \mathbf{h}) + \hat{p}_{ij}(\mathbf{h}_{k+1})(\mathbf{h} - \mathbf{h}_k)}{\mathbf{h}_{k+1} - \mathbf{h}_k}$$

where $\hat{p}_{ij}(\mathbf{h}_k)$ and $\hat{p}_{ij}(\mathbf{h}_{k+1})$ are neighboring transition probability values in an experimental transiogram, and $p_{ij}(\mathbf{h})$ is the interpolated value at the lag $\mathbf{h}$ between $\mathbf{h}_k$ and $\mathbf{h}_{k+1}$. Although the transiogram models obtained by this way are empirical, they still can meet the sum-to-unity constraint in each row of the transiogram matrix, as proved in [15].

*2.1.3 Mathematical model joint modeling*

Modeling of a single transiogram with a mathematical model [12] is similar to modeling of a variogram to some extent, that is, we choose a specific mathematical model to fit the curve of an experimental transiogram by setting the related parameters to some proper values based on the features of the experimental transiogram and/or expert knowledge. The transiogram models obtained by this way are theoretical functions. Based on the mathematical models recommended for variogram modeling [6], a set of mathematical models were suggested for transiogram modeling [12, 26]. In addition, due to the complex features of experimental cross transiograms, a set of gamma distribution-based models were suggested for fitting the peaking feature of some experimental cross transiograms [16]. However, because there is a whole set (i.e., a matrix) of transiograms involved for each categorical field to be simulated, modeling of transiograms has to be done jointly so that the transiogram models obtained can meet the constraints of transition probabilities. By letting one transiogram model to take the left part (i.e., 1.0-others) in each matrix row, the sum-to-unity condition can be met. And by carefully choosing mathematical models and using some tuning, the nonnegativity condition is also easy to be met. Here the "others" in the (1.0-others) model means all other transiogram models in a row of a transiogram model matrix except for the left one. The (1.0-others) model can be calculated by

$$p_{ik}(\mathbf{h}) = 1 - \sum_{\substack{j=1 \\ j \neq k}}^{n} p_{ij}(\mathbf{h})$$

where $n$ is the number of classes in a categorical dataset, $k$ and $j$ denote the head class, and $i$ denotes the tail class of transition probability.

*2.1.4. Mixed joint modeling*

Considering that experimental transiograms involving minor classes may not be reliable due to the very small number of samples for each minor class, linear interpolation may not be an optimal or reasonable choice for obtaining reasonable transiogram models related with minor classes, even though it is highly time efficient. However, the experimental transiograms of major classes may be quite reliable, and consequently linear interpolation can be a good choice not only because of its efficiency but also because of its capability of capturing the complex features of experimental transiograms. Thus, we speculate that a mixed joint modeling method that integrates the merits of both methods (i.e., linear interpolation and mathematical model joint modeling methods) may be a better choice. In this suggested mixed joint fitting method, we use mathematical models to fit experimental transiograms related with minor classes and use linear interpolation to fit experimental transiograms of major classes. To make sure that the transiogram models meet the sum-to-unity constraint, we still use the (1.0-others) model to the left one in each row of the transiogram matrix, with the left one to be normally a major or moderate class.

**2.2 MCRF model and simulation algorithm**

*2.2.1 MCRF model*

The MCRF theory used spatial sequential Bayesian updating to decompose the local conditional probability distribution into likelihood terms and a prior probability term, and then used the proposed spatial conditional independence assumption of nearest neighbors to simplify the multi-point likelihood terms into transition probability terms, so that the model formula can be implemented with transition probabilities, which can be obtained from transiogram models. Suppose we are to estimate the probability of the categorical random variable $Z$ at the location $\mathbf{u}$ being $k$, given the nearest data surrounding $\mathbf{u}$ are $z(\mathbf{u}_1) = l_1, z(\mathbf{u}_2) = l_2, \cdots, z(\mathbf{u}_m) = l_m$, with $\mathbf{h}_1, \mathbf{h}_2, \cdots, \mathbf{h}_m$ being the corresponding distances from $\mathbf{u}$ to $\mathbf{u}_1, \mathbf{u}_2, \cdots, \mathbf{u}_m$, respectively, then the general solution of the conditional probability based on the spatial conditional independence assumption, according to the MCRF theory, can be written as:

$$\Pr(z(\mathbf{u}) = k | z(\mathbf{u}_1) = l_1, \cdots, z(\mathbf{u}_m) = l_m) = \frac{p_{l_1 k}(\mathbf{h}_1) \prod_{i=2}^{m} p_{k l_i}(\mathbf{h}_i)}{\sum_{f=1}^{n} [p_{l_1 f}(\mathbf{h}_1) \prod_{i=2}^{m} p_{f l_i}(\mathbf{h}_i)]}$$

where $n$ is the number of classes, $m$ is the number of surrounding nearest data, and $p_{k l_i}(\mathbf{h}_i)$ is the transition probability from state $k$ to state $l_i$ in $i$th direction with a lag of $\mathbf{h}_i$, which can be fetched from a corresponding transiogram model [11, 19]. Although nearest data are located in different directions, the transiogram models may not have the necessity to consider directions if there is no clear anisotropy. With all of the conditional probabilities being calculated for $k = 1, \ldots, n$, the local conditional probability distribution is obtained as a probability vector of occurrence probabilities of all classes, which can be used for simulating the state (i.e., class label) of the categorical random variable at the location **u**.

Note that due to the quantitative relationship $p_{ij}(\mathbf{h}) = p_{ji}(\mathbf{h})p_j/p_i$ in Markov chain theory for representing reversibility, the above MCRF formula for local condition probability can be written in different forms. For example, we have

$$\frac{p_{l_1k}(\mathbf{h}_1) \prod_{i=2}^{m} p_{kl_i}(\mathbf{h}_i)}{\sum_{f=1}^{n}[p_{l_1f}(\mathbf{h}_1) \prod_{i=2}^{m} p_{fl_i}(\mathbf{h}_i)]} = \frac{p_k \prod_{i=1}^{m} p_{kl_i}(\mathbf{h}_i)}{\sum_{f=1}^{n}[p_f \prod_{i=1}^{m} p_{fl_i}(\mathbf{h}_i)]} = \frac{p_k^{1-m} \prod_{i=1}^{m} p_{l_ik}(\mathbf{h}_i)}{\sum_{f=1}^{n}[p_f^{1-m} \prod_{i=1}^{m} p_{l_if}(\mathbf{h}_i)]},$$

which are mathematically equivalent. When transition probabilities or transiograms are not estimated unidirectionally (e.g., from east to west, from top to bottom), reversibility is automatically assumed.

In practice, we may consider only the four nearest neighbors in four cardinal directions or four quadrants so that the spatial conditional independence assumption can approximately hold; thus the above MCRF model is further simplified to:

$$\Pr(z(\mathbf{u}) = k | z(\mathbf{u}_1) = l_1, \ldots, z(\mathbf{u}_4) = l_4) = \frac{p_{l_1k}(\mathbf{h}_1) \prod_{i=2}^{4} p_{kl_i}(\mathbf{h}_i)}{\sum_{f=1}^{n}[p_{l_1f}(\mathbf{h}_1) \prod_{i=2}^{4} p_{fl_i}(\mathbf{h}_i)]}.$$

In an actual simulation, one may use either a fixed path or a random path, given different modeling tasks or data types.

*2.2.2 MCRF sequential simulation with quadrantal neighborhood*

In this study, the random-path MCRF sequential simulation (MCSS) algorithm [13], which employed a quadrantal neighborhood and whose efficiency was later improved by a spiral neighborhood search [32], is used to conduct stochastic simulations of soil types conditional on irregular sample datasets. The quadrantal neighborhood (i.e., considering one nearest neighbor per quadrant if available within a search window) was proved to be generally optimal for the MCRF approach [19]. Sequential simulation means that the simulated datum in the previous step in a simulation process is always added into the sample dataset for conditioning later simulations at other locations. This strategy is more rational and can make simulated patterns more convergent for categorical variables and smoother for continuous variables. Sequential simulation has been used in geostatistics for a long time, with typical examples of the kriging-based sequential Gaussian simulation and sequential indicator simulation algorithms [6]. In addition, quadrant search (considering a few nearest neighbors per quadrant) was used in kriging long before as a choice for the declustering purpose [9].

## 2.3 Mathematical models used for fitting experimental transiograms

It was found that idealized transiograms generally show four types of shapes: exponential, spherical, Gaussian, and gamma-composite, as displayed in [16] and [27]. Idealized auto-transiograms are usually exponential, while idealized cross-transiograms could be any one of the aforementioned shapes. This knowledge provides some guide to the selection of mathematical models for modeling experimental transiograms. So in this study, when mathematical models are needed for transiogram modeling, we use the exponential model for auto-transiograms and use any of the four model types (i.e., exponential, spherical, Gaussian, and gamma-composite) whenever suitable for cross-transiograms. The functions of these models are as follows:

Exponential model for auto-transiogram:

$$p_{ii}(\mathbf{h}) = 1 - (1-c)[1 - e^{\left(-\frac{3\mathbf{h}}{d}\right)}]$$

Exponential model for cross-transiogram:

$$p_{ij}(\mathbf{h}) = c[1 - e^{\left(-\frac{3\mathbf{h}}{d}\right)}]$$

Gaussian model for cross-transiogram:

$$p_{ij}(\mathbf{h}) = c[1 - e^{\left(-\frac{3\mathbf{h}}{d}\right)^2}]$$

Spherical model for cross-transiogram:

$$p_{ij}(\mathbf{h}) = \begin{cases} c\left[1.5\frac{\mathbf{h}}{d} - 0.5\left(\frac{\mathbf{h}}{d}\right)^3\right], & \text{when } h < d \\ c, & \text{when } h \geq d \end{cases}$$

where $d$ is the range parameter, representing the spatial correlation range, and $c$ is the sill parameter, which may be set to the proportion of the head class. These models could not catch the peak observed in many experimental cross transiograms. To characterize this peaking feature of some cross transiogram models, Li et al. [16] proposed a set of gamma composite models. Gamma distribution is a type of continuous probability distributions with two parameters. The probability density function of Gamma distribution is:

$$f(x|\alpha, \theta) = \frac{1}{\Gamma(\alpha)\theta^\alpha} x^{\alpha-1} e^{-x/\theta}, \alpha, \theta > 0, x \geq 0$$

where $\alpha$ is the shape parameter, and $\theta$ is the scale parameter. $\Gamma(\alpha) = \int_0^\infty t^{\alpha-1}e^{-t}dt$ is the gamma function [5]. The shape of gamma distribution depends on the value of both $\alpha$ and $\theta$. The mean of gamma distribution is $\alpha\theta$, and the variance is $\alpha\theta^2$. When $\alpha \leq 1$, gamma distribution shows a monotonous decreasing trend. When $\alpha > 1$, however, the probability distribution function of gamma distribution shows increasing from zero to a peak and then coming down, which could be used to characterize some cases of cross-transiograms. Combining the gamma function with one of the aforementioned exponential, Gaussian, and spherical model, one can obtain a gamma-based additive composite model. The functions of these cross-transiogram models are as follows:

Gamma-exponential model:

$$p_{ij}(\mathbf{h}) = c\left[1 - e^{\left(-\frac{3\mathbf{h}}{d}\right)} + w\frac{1}{\Gamma(\alpha)\theta^\alpha}\left(\frac{\mathbf{h}}{d}\right)^{\alpha-1} e^{\left(-\frac{\mathbf{h}}{\theta d}\right)}\right], \alpha > 1, \theta > 0$$

Gamma-Gaussian model:

$$p_{ij}(\mathbf{h}) = c\left[1 - e^{\left(-\frac{3\mathbf{h}}{d}\right)^2} + w\frac{1}{\Gamma(\alpha)\theta^\alpha}\left(\frac{\mathbf{h}}{d}\right)^{\alpha-1} e^{\left(-\frac{\mathbf{h}}{\theta d}\right)}\right], \alpha > 1, \theta > 0$$

Gamma-spherical model:

$$p_{ij}(\mathbf{h}) = \begin{cases} c\left[1.5\frac{\mathbf{h}}{d} - 0.5\left(\frac{\mathbf{h}}{d}\right)^3 + w\frac{1}{\Gamma(\alpha)\theta^\alpha}\left(\frac{\mathbf{h}}{d}\right)^{\alpha-1} e^{\left(-\frac{\mathbf{h}}{\theta d}\right)}\right], & \text{when } h < d, \alpha > 1, \theta > 0 \\ c\left[1 + w\frac{1}{\Gamma(\alpha)\theta^\alpha}\left(\frac{\mathbf{h}}{d}\right)^{\alpha-1} e^{\left(-\frac{\mathbf{h}}{\theta d}\right)}\right], & \text{when } h \geq d, \alpha > 1, \theta > 0 \end{cases}$$

where $d$ is the range parameter of the matching model (i.e., the exponential, Gaussian, or spherical model), $\alpha$ is the shape parameter of the gamma distribution, $\theta$ is the scale parameter of the gamma distribution, $w$ is a weight parameter for users to adjust how much of the gamma distribution is used in the composite model, and $c$ is the sill parameter, which may be set to the proportion of the head class $j$ [16].

## 3. Data for case studies

This study aims to examine the sensitivity of the MCRF model to the transiogram models obtained using the three different ways - linear interpolation, mathematical model joint fitting, and mixed joint fitting and test the impacts of these transiogram joint modeling methods on simulation accuracy, especially the accuracy of minor class, with two case studies. Case study one used a dataset downloaded from Natural Resources Conservation Service (NRCS), US Department of Agriculture. This dataset was previously used in [13] for testing the MCSS algorithm. The soil classification map covers a 3.5 km × 2.56 km area in Iowa County, Wisconsin. The soil series in the map were combined into seven classes: (1) loamy alluvium, (2) Sparta, (3) Plainfield, (4) Dakota, (5) Richwood, (6) others, and (7) peat and muck, among which class 5 is a typical minor class (accounting for 1.94% of total area). The study area was discretized into a grid of 175 × 128 pixels using a pixel size of 20m × 20m. Two sample datasets, a dense one (646 points, accounting for 2.9% of total pixels) and a sparse one (179 points, accounting for 0.8% of total pixels), were randomly taken from the original raster map data as the sample data input to conditional simulations and also for estimating experimental transiograms. Fig. 1 shows the original reference soil map and sample datasets.

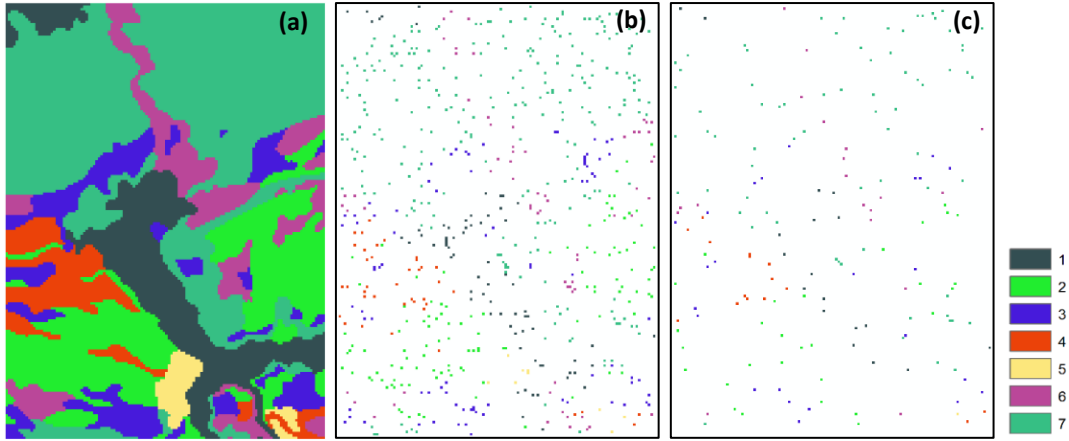

**Fig. 1** The reference raster soil type map (adapted from a subarea of the Iowa County soil map) (a) and the dense and the sparse random point sample datasets (b and c) taken from the map for Case study one.

Case study two used a soil textural class dataset obtained from NRCS. With this example, we aim to examine the sensitivity of simulated results to different transiogram joint modeling methods in a relatively larger area. The study area for this case is Scotland, a town of Windham County in eastern Connecticut, with an area of 48.4 km$^2$. In the original dataset there are 11 soil textural classes. Because several textural classes (e.g. mucky peat, loam, and loamy fine sand) are too minor for transiogram model estimation, and also for simplicity of modeling, we regrouped the 11 textural classes into 6 classes. The 6 classes are (1) sandy loam and fine sand, (2) moderately to highly decomposed plant material, (3) water surface, (4) muck and mucky peat, (5) silt loam, and (6) slightly decomposed plant material, among which classes 3, 5, and 4 are minor classes (accounting for 0.43%, 1.74%, and 2.98% of total area, respectively). The study area was discretized into a grid of 235 × 415 pixels using a pixel size of 25m × 25m. A total of 1590 random sample data (2.04% of the total pixels) were retrieved from the raster map for estimating experimental transiograms and performing conditional simulation (Fig. 2).

In both case studies, when mathematical models were used for transiogram modeling, the sills of transiogram models were set to the proportions of the corresponding head classes in sample data. The rationale behind this is that the sills of idealized transiograms are theoretically equal to and actually approach the proportions of the corresponding head classes [12, 27]. By this way, it is also easier to make the transiogram models in each matrix row meet the sum-to-unity constraint. However, there is an exception for minor class 5 in the sparse sample dataset in Case study one. When mathematical model joint fitting was performed, we set the sills of its transiogram models to its proportion in the dense sample dataset. This is because this class was severely under-sampled and has only one sample point in the sparse sample dataset, which makes it not able to generate meaningful experimental transiograms.

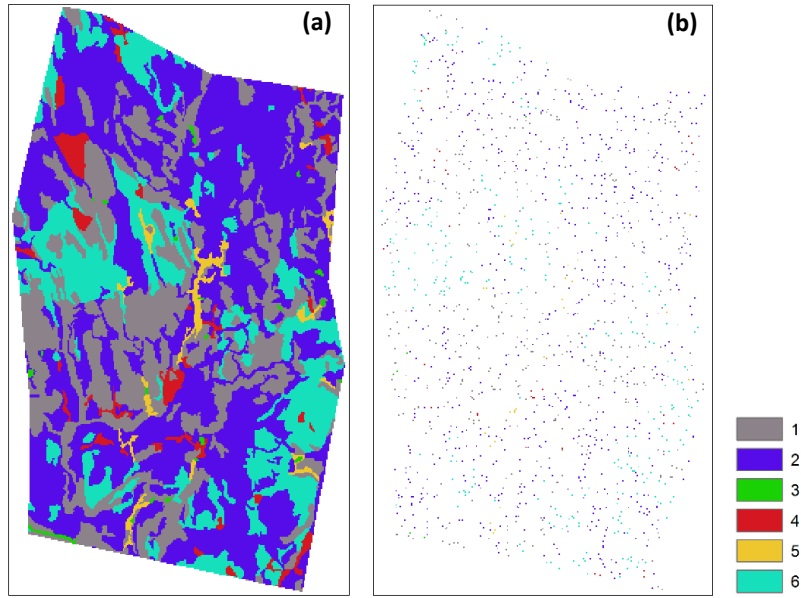

**Fig. 2** The reference surface soil texture class map (adapted from the Scotland soil textural map of Windham County) (a) and the random point sample dataset (b) taken from the map for Case study two.

## 4. Results and analysis

### 4.1 Transiogram joint modeling

Omni-directional experimental transiograms were estimated from all sample datasets for the two case studies, because we did not consider the anisotropy of soil type patterns due to the fact that there are not identical anisotropic directions in both of the reference soil maps. After repeated testing, a tolerance width of 3 pixel length was used for the dense sample dataset and a tolerance of 7 pixel length was used for the sparse sample dataset in Case study one; a tolerance width of 4 pixel length was used for the only sample dataset in Case study two. Using the three transiogram joint modeling methods, we obtained the transiogram models for each sample datasets.

*4.1.1 Transiogram models for Case study one*

Tables 1, 2, and 3 show the parameters for characterizing the transiogram models with tail class 1 in the dense sample dataset, tail class 4 in the sparse sample dataset, and tail class 5 in the sparse sample dataset in Iowa County dataset, respectively; and Figures 3, 4, and 5 plot out the corresponding subsets of the interpolated and mathematical model fitted transiograms. The *x*-axis of the plotted transiograms is the lag distance in pixel length, and *y*-axis is the transition probability. While the theoretical transiogram models (by mathematical model joint fitting) and the empirical transiogram models (by linear interpolation) have some common features, they also have some differences: (1) The former capture the change trend of the latter, that is, they are mostly able to represent the experimental transiograms; (2) while the latter may contain some noise, they also have some features that are not captured by the theoretical models.

**Table 1** Parameters used in mathematical model-fitted transiogram models with tail class 1, based on the Iowa dense sample dataset for Case study one.

|  | Model type | Sill [a] | Range | $\alpha$ | $\theta$ | Weight |
|---|---|---|---|---|---|---|
| $p_{11}(h)$ | Exponential | 0.1115 | 40 | | | |
| $p_{12}(h)$ | Gamma-exponential | 0.1765 | 80 | 4.0 | 0.3 | 1.4 |
| $p_{13}(h)$ | Gamma-spherical | 0.1269 | 40 | 2.5 | 0.75 | 0.6 |
| $p_{14}(h)$ | Gamma-Gaussian | 0.0604 | 25 | 1.8 | 1.0 | 1.5 |
| $p_{15}(h)$ | Gamma-exponential | 0.0139 | 22 | 2.0 | 0.4 | 3.0 |
| $p_{16}(h)$ | Exponential | 0.1022 | 40 | | | |
| $p_{17}(h)$ | 1.0 – others [b] | 0.4086 | | | | |

[a] The sill values are the same as the head class proportions

[b] "others" means the sum of transiogram models with other head classes

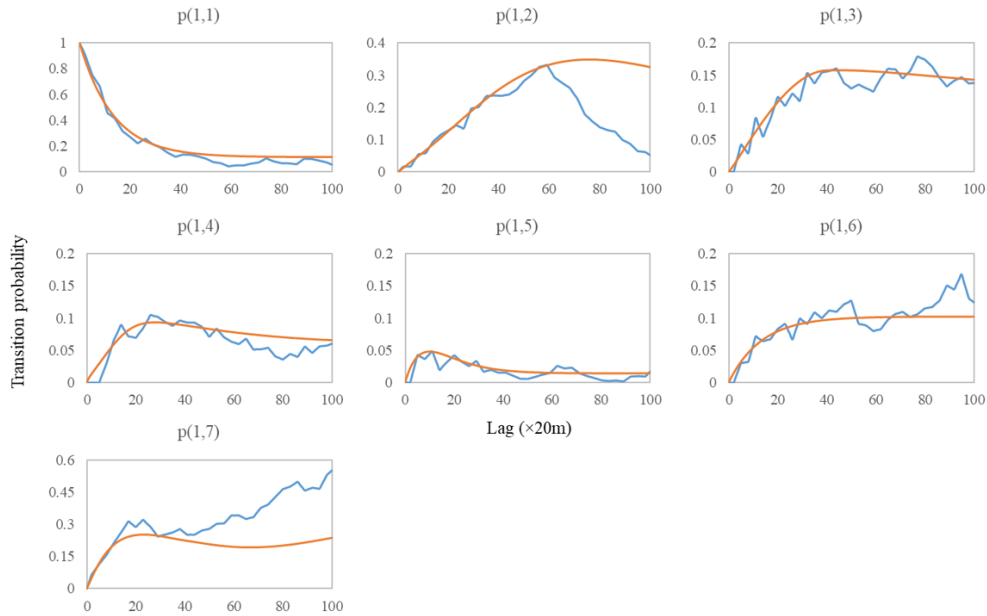

**Fig. 3** Transiogram models from linear interpolation (blue) and mathematical model joint fitting (red) with tail class 1 about the Iowa dense sample dataset for Case study one. Here p(i,j) means $p_{ij}(h)$.

Auto-transiograms ($p_{11}(h)$ in Fig. 3 and $p_{44}(h)$ in Fig. 4) show approximate exponential shape starting from point (0, 1) and gradually decreasing to their sills (the sill of each theoretical model is set to the proportion of the head class). We found that many cross-transiograms are better fitted with the gamma composite models because they first increase to a peak and then gradually decrease to a stable level (usually their sills). This happens when two classes tend to be neighboring classes. For example, $p_{14}(h)$ in Fig. 3 and $p_{41}(h)$ in Fig. 4 show this peaking feature because class 1 and class 4 often appear as neighboring classes in the original map. $p_{12}(h)$ in Fig. 3 reaches a very high peak 0.3322 at the lag of 59 while its sill should be 0.1765, for which the reason is the juxtaposition tendency of class 1 and class 2.

**Table 2** Parameters used in mathematical model-fitted transiogram models with tail class 4, based on the Iowa sparse sample dataset for Case study one

|  | Model type | Sill [a] | Range | $\alpha$ | $\theta$ | Weight |
|---|---|---|---|---|---|---|
| $p_{41}(h)$ | Gamma-exponential | 0.1006 | 50 | 2.5 | 0.5 | 1.5 |
| $p_{42}(h)$ | Gamma-exponential | 0.1620 | 30 | 2.0 | 1.5 | 2.0 |
| $p_{43}(h)$ | Gamma-exponential | 0.1341 | 20 | 2.5 | 0.75 | 1.0 |
| $p_{44}(h)$ | Exponential | 0.0838 | 36 |  |  |  |
| $p_{45}(h)$ | Gamma-exponential | 0.0139 | 10 | 2.0 | 0.5 | 4.0 |
| $p_{46}(h)$ | Exponential | 0.1061 | 36 |  |  |  |
| $p_{47}(h)$ | 1.0 – others [b] | 0.3995 |  |  |  |  |

[a] The sill values are the same as the head class proportions except for $p_{45}(h)$

[b] "others" means the sum of transiogram models with other head classes

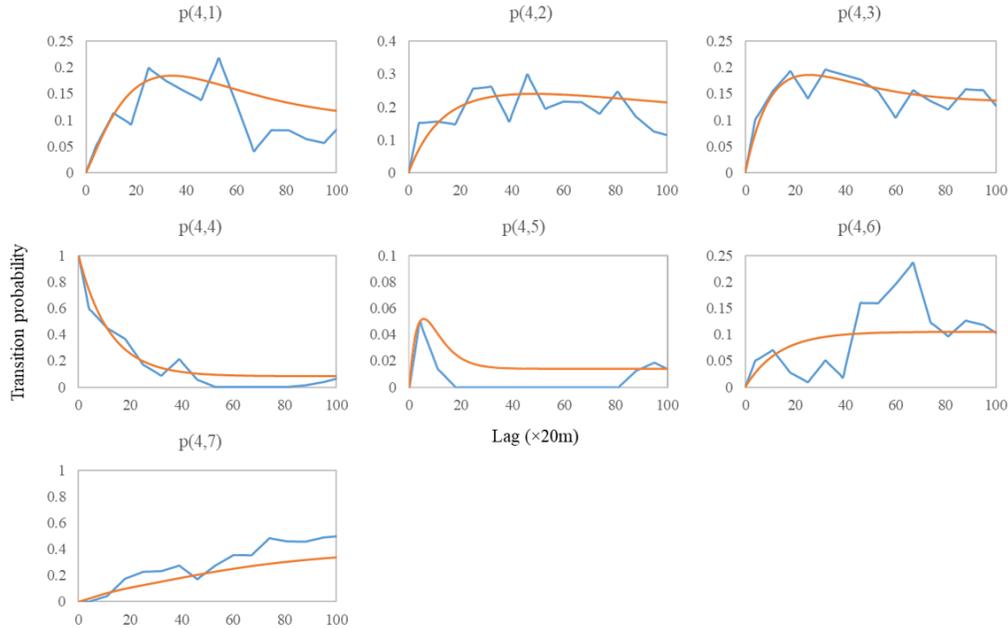

**Fig. 4** Transiogram models from linear interpolation (blue) and mathematical model joint fitting (red) with tail class 4 about the Iowa sparse sample dataset for Case study one. Here p(i,j) means $p_{ij}(h)$.

Experimental cross-transiograms tend to have more complex shapes compared with experimental auto-transiograms. Thus, the theoretical cross-transiogram models might not be able to capture all the main features of experimental transiograms (e.g., $p_{12}(h)$ in Fig. 3 and $p_{46}(h)$ in Fig. 4). In this study, we set the parameters of the fitted mathematical models to proper values by emphasizing the well fit of the low lag section of experimental transiograms. This is because usually only the transition probabilities at low lag values are used by the MCRF model in simulation. Therefore, the low lag sections of transiogram models have higher impact on simulation accuracies.

The sample data of class 4 accounts for 6.04% of all samples in the dense sample dataset and 8.38% in the sparse sample dataset, and that of class 5 accounts for 1.39% of all samples in the dense sample dataset and 0.56% in the sparse sample dataset (actually class 5 is strongly under-sampled and has only one sample point in the sparse sample dataset). Both are relatively minor classes compared with other 5 classes. Therefore, when generating the transiogram models using the mixed joint modeling method, we used linear interpolation for transiograms with large head and tail classes (1, 2, 3, and 6), and used mathematical models for transiograms with minor head or tail classes (4 and 5). The (1.0-others) model is used for the transiograms with head class 7 (a major class) in all transiogram modeling methods related with mathematical models. In addition, when mathematical model fitting is performed, sills for transiogram models with head class 5 were set to 0.0139 (same as the proportion of class 5 in the dense sample dataset) in order to generate reliable transiogram models (see $p_{45}(h)$ in Fig. 4).

Because class 5 only has one point in the sparse sample dataset, there is actually no way to get the experimental auto-transiogram of class 5; therefore, we set the range parameter of its auto-transiogram model to its corresponding value for the dense sample dataset in order to obtain a reliable auto-transiogram model by incorporating expert knowledge (Table 3). Apparently, experimental transiograms with class 5 estimated from the sparse sample dataset are very unreliable with abnormal features; that is why the theoretical models, which are actually based on the theoretical models for the dense sample dataset in terms of model choice, do not have a well fit (Fig. 5).

**Table 3** Parameters used in mathematical model-fitted transiogram models with tail class 5, based on the Iowa sparse sample dataset for Case study one

|  | Model type | Sill [a] | Range | $\alpha$ | $\theta$ | Weight |
|---|---|---|---|---|---|---|
| $p_{51}(h)$ | Gamma-Gaussian | 0.1006 | 40 | 2.5 | 0.5 | 2.0 |
| $p_{52}(h)$ | Gamma- Gaussian | 0.162 | 40 | 5.0 | 0.5 | 3.0 |
| $p_{53}(h)$ | Gamma-exponential | 0.1341 | 4 | 6.0 | 0.6 | 8.0 |
| $p_{54}(h)$ | Exponential | 0.0838 | 10 | | | |
| $p_{55}(h)$ | Exponential | 0.0139 | 18 | | | |
| $p_{56}(h)$ | Gamma-Gaussian | 0.1061 | 25 | 3.0 | 0.5 | 2.0 |
| $p_{57}(h)$ | 1.0 – others [b] | 0.3995 | | | | |

[a] The sill values are the same as the head class proportions except for $p_{54}(h)$ and $p_{55}(h)$

[b] "others" means the sum of transiogram models with other head classes

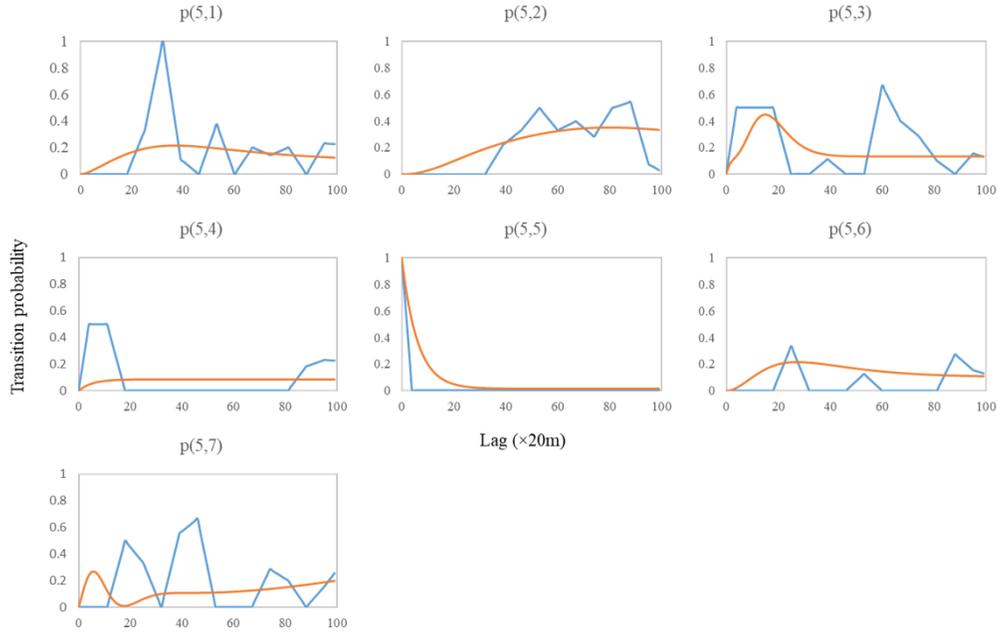

**Fig. 5** Transiogram models from linear interpolation (blue) and mathematical model joint fitting (red) with tail class 5 about the Iowa sparse sample dataset for Case study one. Because class 5 has only one sample datum, these experimental transiograms are unreliable with abnormal features. Here p(i,j) means $p_{ij}$(h).

### 4.1.2 Transiogram models for Case study two

Table 4 shows the parameters for characterizing the transiograms with tail class 2 in the Scotland dataset, and Figure 6 plots the corresponding transiogram models generated by the linear interpolation and mathematical model fitting methods. One can see that the auto-transiogram in this subset (i.e., $p_{22}$(h)) shows an exponential shape. However, the cross-transiograms have no apparent peak features; instead, their shapes also tend to be exponential, gradually increasing to their sills (i.e., proportions of corresponding head classes). The reason may be that class 2 does not have the preference for its neighboring classes. It also can be seen that class 2 has an auto-correlation range of about 20 pixel length, which is larger than its cross-correlation ranges with other classes. This means that the mean patch size of class 2 is relatively large. For cross-transiograms, the correlation ranges reflect the interdependence of every class pairs. In this case, it seems that the cross-correlation ranges between class 2 (a major class, accounting for 47% of all sample data) and other classes (smaller and minor classes) are consistent with the proportions of other classes. For example, class 3 and class 5 are very small in proportion, and their cross-correlation ranges with class 2 are much smaller than those of other classes with class 2.

Similar to Case study one, three different sets of transiogram models were obtained for performing simulations in Case study two. Because classes 3, 4, and 5 all have a small number of samples (0.44%, 2.77% and 1.45%, respectively), when using the mixed joint modeling method to generate transiogram models, we use the linear interpolation method when the head and tail classes are major classes (i.e., 1 or 2), and use the mathematical model joint modeling method when the head or tail class is a minor class (i.e., 3, 4, or 5). The (1.0-others) method was used when the head class is 6 (a moderate class).

**Table 4** Parameters used in mathematical model-fitted transiogram models with tail class 2, based on the Scotland sample dataset for Case study two.

|         | Model type     | Sill [a]  | Range |
|---------|----------------|-----------|-------|
| $p_{21}(h)$ | exponential    | 0.2931    | 15    |
| $p_{22}(h)$ | exponential    | 0.4704    | 20    |
| $p_{23}(h)$ | exponential    | 0.0044    | 3     |
| $p_{24}(h)$ | exponential    | 0.0277    | 10    |
| $p_{25}(h)$ | spherical      | 0.0144    | 6     |
| $p_{26}(h)$ | 1.0 – others [b] | 0.1899  |       |

[a] The sill values are the same as the head class proportions

[b] "others" means the sum of transiogram models with other head classes

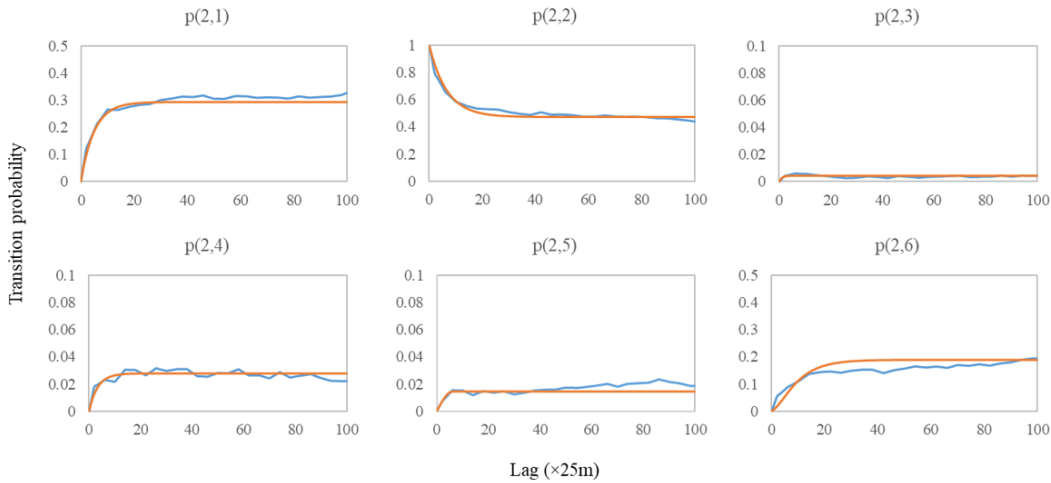

**Fig. 6** Transiogram models from linear interpolation (blue) and mathematical model joint fitting (red) with tail class 2 about the Scotland dataset for Case study two. Here p(i,j) means $p_{ij}(h)$.

### 4.2 Simulated results of Case study one

*4.2.1 Realization outputs*

Figure 7 shows examples of simulated realizations by MCSS algorithm conditioned on the Iowa dense and sparse sample datasets using the three transiogram joint modeling methods. The simulation search radiuses are 30 and 50 pixel lengths for the dense and sparse sample datasets, respectively. All six realizations shown here were generated using the same random number seed. Compared with the reference map (Fig. 1), one can see that the realization maps based on both the dense and sparse sample datasets reproduced the general spatial pattern of the seven classes. However, fine linear features were not reproduced due to the fact that their complexity cannot be effectively captured by two-point statistics (here transiograms) and pure sample data. Table 5 shows the averaged simulation accuracies (percentages of correctly simulated pixels) of realization maps and individual classes. There are differences in simulation accuracy between realizations, especially between some classes, based on different sets of transiogram models and the same sample dataset. Minor classes (e.g., 3 and 5) have relatively lower accuracy than major classes (e.g., 1 and 7).

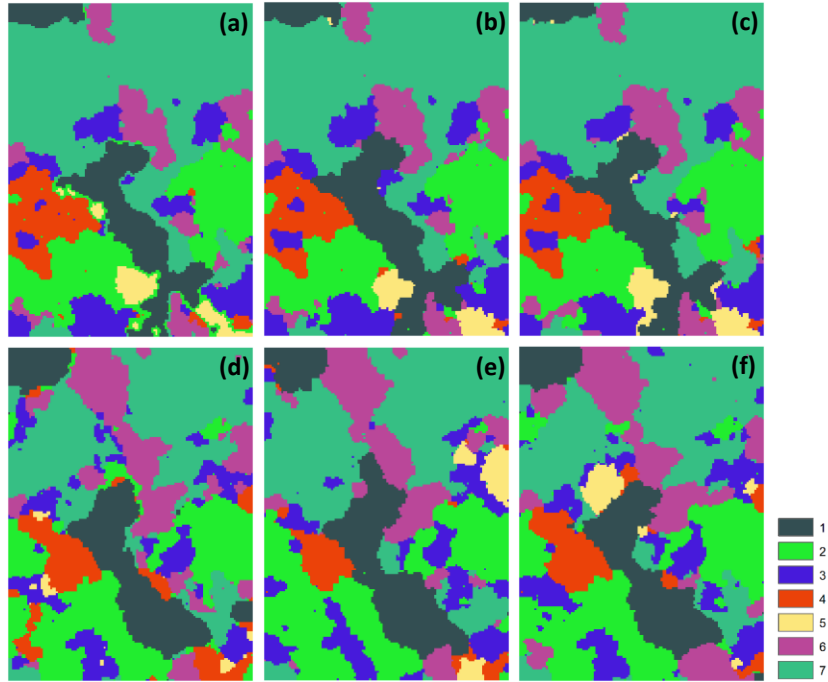

**Fig. 7** Simulated realizations by the MCSS algorithm for Case study one. (a), (b), and (c) are examples of simulated realization maps based on the dense sample dataset, with linear interpolation, mathematical model, and mixed joint modeling methods, respectively. (d), (e), and (f) are examples of simulated realization maps based on the sparse sample dataset, with linear interpolation, mathematical model, and mixed joint modeling methods, respectively.

**Table 5** Averaged accuracies (percentages of correctly simulated pixels) of simulated realizations and single classes using three different transiogram joint modeling methods for Case study one (averaged from all simulated realizations).

| Sample dataset | Transiogram modeling method | Overall | Class 1 | Class 2 | Class 3 | Class 4 | Class 5 | Class 6 | Class 7 |
|---|---|---|---|---|---|---|---|---|---|
| Dense | Linear interpolation | 76.38 | 73.33 | 75.50 | 64.59 | 70.53 | 56.28 | 61.85 | 85.97 |
|  | Mathematical model | 76.65 | 80.06 | 73.02 | 66.64 | 68.38 | 73.14 | 65.66 | 83.77 |
|  | Mixed | 76.99 | 75.68 | 74.35 | 63.42 | 68.22 | 74.99 | 66.09 | 86.00 |
| Sparse | Linear interpolation | 63.88 | 68.02 | 62.47 | 33.73 | 57.71 | 6.02 | 44.95 | 78.79 |
|  | Mathematical model | 64.68 | 67.50 | 63.11 | 40.11 | 56.72 | 7.15 | 49.79 | 78.02 |
|  | Mixed | 64.10 | 68.74 | 64.64 | 37.38 | 60.32 | 7.98 | 50.18 | 75.46 |

One purpose of this study is to test whether incorporating expert knowledge could improve the simulation performance for minor classes. For the simulations based on the dense sample dataset, the averaged
18

simulation accuracies of the smallest class 5 using the mathematical model fitting method and the mixed method have obvious improvement over that using the linear interpolation method for transiogram modeling (Table 5). However, the overall simulation accuracies using different sets of transiogram models agree with each other quite well. Note that the smallest class – class 5 in the realizations based on the sparse sample dataset has very low averaged accuracies (6.02% to 7.98%) for all of the three transiogram joint modeling methods. The main reasons should be that class 5 is severely under-sampled (only one sample point) in the sparse sample dataset and consequently its occurrence in simulated realizations is very random.

**Table 6** Proportions (%) of individual classes in the reference map, dense and sparse random sample datasets, and the averaged proportions of individual classes in simulated realizations for Case study one (the Iowa dataset)

|  |  | Class 1 | Class 2 | Class 3 | Class 4 | Class 5 | Class 6 | Class 7 |
|---|---|---|---|---|---|---|---|---|
| Reference map | | 12.84 | 18.75 | 9.78 | 6.31 | 1.94 | 9.80 | 40.58 |
| Dense sample dataset | Sample data | 11.15 | 17.65 | 12.69 | 6.04 | 1.39 | 10.22 | 40.86 |
| | Linear interpolation | 11.73 | 19.12 | 10.25 | 6.57 | 2.00 | 10.12 | 40.21 |
| | Mathematical model | 13.40 | 16.69 | 11.34 | 6.32 | 2.77 | 11.25 | 38.23 |
| | Mixed | 12.17 | 17.13 | 10.05 | 6.34 | 3.18 | 11.28 | 39.85 |
| Sparse sample dataset | Sample data | 10.06 | 16.20 | 13.41 | 8.38 | 0.56 | 10.61 | 40.78 |
| | Linear interpolation | 12.96 | 19.64 | 7.76 | 7.20 | 1.06 | 10.84 | 40.54 |
| | Mathematical model | 12.86 | 19.06 | 9.78 | 6.35 | 0.56 | 12.60 | 38.79 |
| | Mixed | 13.28 | 20.13 | 8.49 | 6.73 | 0.59 | 12.69 | 38.09 |

Table 6 shows the proportions of individual classes in the sample datasets and the averaged proportions of individual classes in simulated realizations (averaged from 100 realizations). The averaged proportions of different classes in realizations are close to their proportions in corresponding sample datasets. While some classes show some overestimation in class proportion, some other classes show some underestimation. This is understandable because both sample data and transiogram models control the outputs of simulations. However, the transiogram models based on experimental transiograms and expert knowledge may deviate more or less from the exact spatial auto and cross correlations of classes. This is especially true for the empirical transiogram models (by linear interpolation of experimental transiograms) based on a very sparse sample dataset with a relatively large tolerance width. In addition, subjectivity in setting the parameters (i.e., choosing model type and correlation range) of theoretical transiogram models (by mathematical model fitting) is also inevitable. Therefore, it is not easy to accurately reproduce the class proportions of sample data in simulated realizations.

*4.2.2 Optimal outputs*

Optimal output maps were generated based on the largest class occurrence probabilities at each location, which are estimated from a number of simulated realizations. Fig. 8 shows the simulated optimal maps of the Iowa case study area using different transiogram modeling methods. The optimal maps conditioned on the dense sample dataset well captured the general spatial pattern of classes, similar to that of the reference soil map except for the fine linear features. The optimal maps conditioned on the sparse sample dataset yielded relatively simpler patterns with lower accuracy due to less conditioning samples, but they still captured the major soil pattern in the original reference soil map.

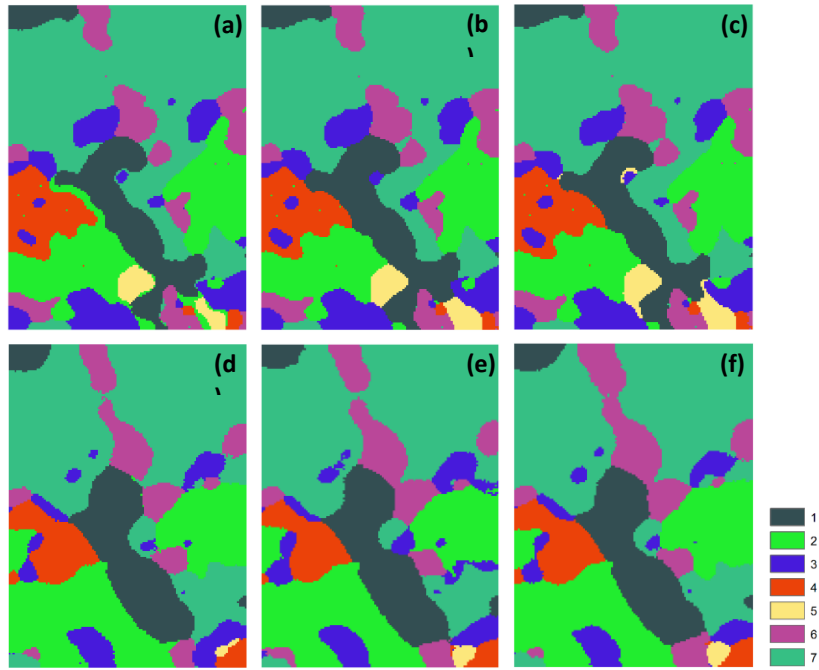

**Fig. 8** Optimal output maps generated by the MCSS algorithm for Case study one. Maps in first row are based on the dense sample dataset, with (a) linear interpolation, (b) mathematical model, and (c) mixed transiogram joint modeling methods, respectively. Maps in second row are based on the sparse sample dataset, with (d) linear interpolation, (e) mathematical model, and (f) mixed transiogram joint modeling methods, respectively.

Minor differences between the optimal maps based on the same sample dataset can be observed in the center and the lower right areas of the maps, where the interclass relationships are more complex compared with those in other areas. These minor differences are mainly caused by the differences between the transiogram models used for the corresponding simulations. It is apparent that the spatial patterns of the optimal maps conditioned on the sparse sample dataset are simpler than those based on the dense sample dataset. However, the spatial patterns of classes in the optimal output maps based on the same sample dataset are basically consistent with each other, although they used different sets of transiogram models.

Because of the smoothing effect of optimal output maps [13], one can see that the optimal maps (Fig. 8) have more smooth boundaries, less complex shapes, and less small scattered patches compared with the corresponding realization maps (Fig. 7). It should be noted that one might easily get an illusion here that the smallest class 5 was strongly underestimated in optimal maps and simulated realizations based on the sparse sample dataset, compared with those maps based on the dense sample dataset. This is not the case as shown in [11] and [29], and the reason for this problem at here is mainly the severe under-sampling of the minor class 5 in the sparse sample dataset. An under-sampled class should be underestimated because the estimation is based on the sample data.

**Table 7**. Accuracies (percentages of correctly simulated pixels) of optimal maps and individual classes using three different transiogram joint modeling methods in Case study one (the Iowa dataset) (corresponding to Fig. 8).

| Sample dataset | Transiogram modeling method | Overall | Class 1 | Class 2 | Class 3 | Class 4 | Class 5 | Class 6 | Class 7 |
|---|---|---|---|---|---|---|---|---|---|
| Dense | Linear interpolation | 79.80 | 79.17 | 79.47 | 67.73 | 76.29 | 63.82 | 65.39 | 87.86 |
|  | Mathematical model | 79.79 | 83.94 | 76.71 | 69.69 | 74.45 | 74.88 | 69.26 | 85.94 |
|  | Mixed | 80.13 | 79.94 | 78.14 | 66.82 | 73.74 | 79.49 | 69.63 | 87.89 |
| Sparse | Linear interpolation | 71.17 | 73.16 | 75.57 | 38.48 | 66.45 | 5.76 | 53.64 | 84.49 |
|  | Mathematical model | 70.85 | 73.85 | 72.78 | 44.18 | 61.71 | 13.59 | 58.93 | 82.48 |
|  | Mixed | 71.08 | 74.93 | 76.45 | 39.62 | 66.45 | 14.75 | 60.11 | 81.03 |

Table 7 shows the simulation accuracies of the optimal output maps and individual classes in the maps. Compared with the averaged accuracies of realizations shown in Table 5, the accuracies of optimal maps are 3% - 6% higher for the simulations conditioned on the dense sample dataset, and more than 6% higher for the simulations conditioned on the sparse sample dataset.

One can see that the optimal output maps based on the three different transiograms joint modeling methods yielded similar overall accuracies and mostly similar class accuracies for each sample dataset. These results mean that the three methods of obtaining transiogram models perform similarly in terms of simulation optimal outputs. However, the mathematical model fitting method and the mixed method for transiogram joint modeling indeed made improvement in accuracy for the smallest class 5 over the linear interpolation method (by 11.06% and 15.67%, respectively, based on the dense sample dataset, and by 7.83% and 8.99%, respectively, based on the sparse sample dataset). This means that using mathematical model-fitted transiogram models is helpful in dealing with the unreliability problems of experimental transiograms of the minor class. In addition, because class 5 only has one sample point in the sparse sample dataset (severely under-sampled), its accuracies in the optimal maps based on the sparse sample dataset are extremely low due to its strongly random occurrence. Smaller classes tend to have lower accuracies in stochastic simulations because their spatial occurrences tend to be more random.

**Table 8** Proportions of individual classes in the original reference map, random sample datasets, and optimal output maps for Case study one (the Iowa dataset).

|  |  | Class 1 | Class 2 | Class 3 | Class 4 | Class 5 | Class 6 | Class 7 |
|---|---|---|---|---|---|---|---|---|
| Reference map |  | 12.84 | 18.75 | 9.78 | 6.31 | 1.94 | 9.80 | 40.58 |
| Dense sample dataset | Sample data | 11.15 | 17.65 | 12.69 | 6.04 | 1.39 | 10.22 | 40.86 |
|  | Linear interpolation | 12.35 | 18.84 | 9.99 | 6.67 | 1.83 | 9.78 | 40.54 |
|  | Mathematical model | 13.67 | 16.84 | 11.13 | 6.42 | 2.45 | 10.53 | 38.96 |
|  | Mixed | 12.51 | 17.47 | 9.90 | 6.41 | 2.77 | 10.57 | 40.37 |
| Sparse sample dataset | Sample data | 10.06 | 16.20 | 13.41 | 8.38 | 0.56 | 10.61 | 40.78 |
|  | Linear interpolation | 13.31 | 22.21 | 6.36 | 5.89 | 0.22 | 9.17 | 42.84 |
|  | Mathematical model | 13.47 | 20.95 | 8.42 | 5.47 | 0.53 | 10.75 | 40.41 |
|  | Mixed | 13.71 | 22.27 | 5.88 | 6.01 | 0.58 | 11.66 | 39.89 |

Because of the smoothing effect, class proportions in optimal maps usually may have some deviation from their proportions in the sample dataset, and minor classes tend to be underrepresented to some extent, particularly when sample data for conditioning are sparse, because random occurrences with low occurrence probabilities are smoothed out in optimal maps. Table 8 shows the proportions of individual classes in the sample datasets and the optimal output maps for Case study one. It can be seen that class proportions in the optimal maps indeed have some deviation from that in the corresponding sample datasets. However, minor class underrepresentation mainly occur for classes 3 and 4 in the optimal maps based on the sparse sample dataset, but does not occur for the smallest class 5, except in the optimal map based on the sparse sample dataset and the linear interpolation transiogram modeling method. On the contrary, it seems that the transiogram models from the mathematical model fitting cause some overrepresentation of the minor class 5.

### 4.3 Simulated results of Case study two

*4.3.1 Realization outputs*

Figure 9 shows examples of simulated realizations by the MCSS algorithm for Case study two (i.e., the Scotland dataset) using the three transiogram joint modeling methods, with the same random number seed. The simulation search radius is 30 pixel lengths. It can be seen that the general spatial pattern of the reference map was well reproduced. However, the directional trends of some patches (mainly some patches of class 1 and class 6) were not captured in the realization maps due to the inability of two-point statistics (here transiograms) and pure sample data in capturing complex linear features. Table 9 shows the averaged simulation accuracy of the realization maps and individual classes for this case study. The averaged accuracies using different sets of transiogram models here are quite consistent with each other. Using mathematical models and expert knowledge for transiogram modeling does not show effect in improving the simulation accuracy of minor classes (e.g., classes 3, 4 and 5), because the experimental transiograms estimated from the sample dataset are reliable. Table 10 shows the proportions of individual classes in the reference map, sample dataset, and realization maps. The simulated results using the three different sets of transiogram models were mostly consistent among realizations and similar to their corresponding values in the sample dataset, but it seems that there exists to some extent a tendency of major class being underrepresented and minor class being overrepresented.

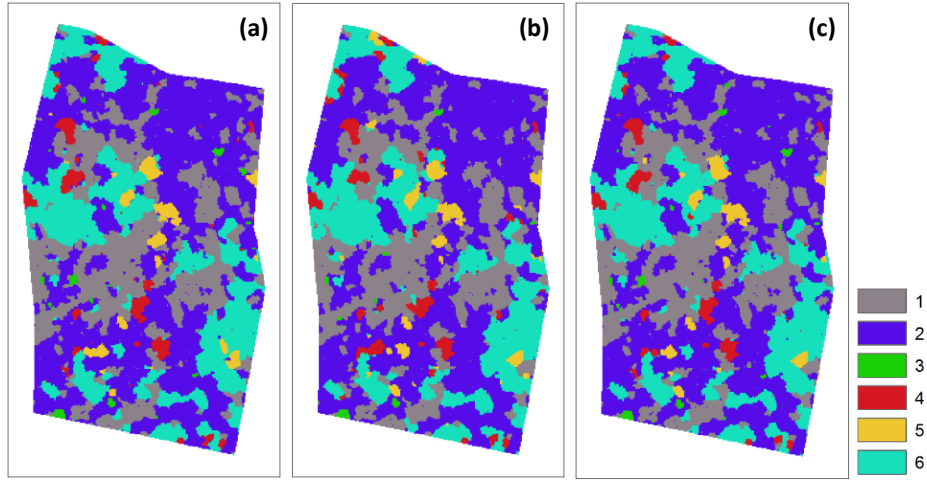

**Fig. 9** Simulated realizations by the MCSS algorithm for the Scotland case study area. (a), (b), and (c) are the examples of simulation realization maps with linear interpolation, mathematical model, and mixed transiogram joint modeling methods, respectively.

**Table 9** Averaged simulation accuracies (%) of realizations and individual classes using the three different transiogram joint modeling methods for Case study two (the Scotland dataset).

| Transiogram Modeling method | Overall | Class 1 | Class 2 | Class 3 | Class 4 | Class 5 | Class 6 |
| --- | --- | --- | --- | --- | --- | --- | --- |
| Linear interpolation | 65.33 | 60.77 | 68.30 | 16.11 | 46.38 | 31.01 | 71.93 |
| Mathematical model | 65.50 | 60.69 | 68.27 | 14.33 | 47.51 | 29.69 | 73.06 |
| Mixed | 65.48 | 61.02 | 68.71 | 15.25 | 47.27 | 30.06 | 71.29 |

**Table 10** Proportions (%) of individual classes in the reference map, random sample dataset, and the averaged proportions of individual classes in realization maps for Case study two (the Scotland dataset)

|  | Class 1 | Class 2 | Class 3 | Class 4 | Class 5 | Class 6 |
| --- | --- | --- | --- | --- | --- | --- |
| Reference map | 28.19 | 48.05 | 0.43 | 2.98 | 1.74 | 18.61 |
| Sample data | 29.31 | 47.04 | 0.45 | 2.77 | 1.44 | 18.99 |
| Linear interpolation | 27.56 | 43.55 | 0.65 | 3.61 | 2.21 | 22.42 |
| Mathematical model | 27.26 | 43.26 | 0.53 | 3.73 | 2.35 | 22.87 |
| Mixed | 27.69 | 43.88 | 0.63 | 3.53 | 2.17 | 22.10 |

*4.3.2 Optimal outputs*

Figure 10 shows the optimal output maps based on the Scotland sample dataset. Overall, three optimal maps captured the soil spatial patterns. However, the soil patches in the optimal maps tend to be circular in shape and have round boundaries, and the vertical direction patterns of classes 1 and 5 in the right part of the study area are not captured.

These insufficiencies are normal when simulation is purely based on sample data because of the smoothing effect, the use of omni-directional transiograms, and the inability of two-point statistics.

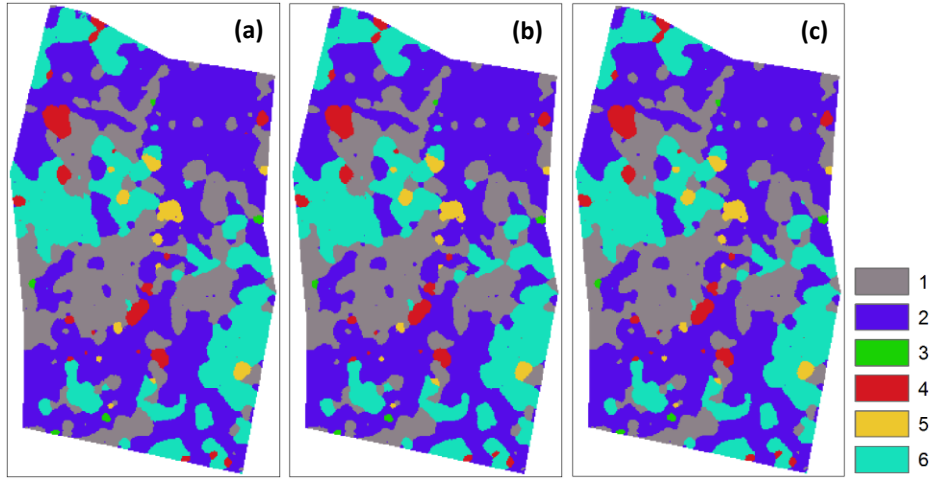

**Fig. 10** Optimal output maps generated by the MCSS algorithm for the Scotland case study, using the transiogram models generated by (a) linear interpolation, (b) mathematical model, and (c) mixed transiogram joint modeling methods, respectively.

Tables 11 and 12 show the simulation accuracies and class proportions of the optimal maps, respectively, in this case study. Since the differences between the transiogram model sets generated by the three different methods are not obvious (see Fig. 6), similar results were obtained in optimal maps in terms of simulation accuracy. It is expected that the minor class 3 has a very low accuracy due to its small number of samples, which causes stronger spatial uncertainty compared with large classes. The proportions of individual classes in optimal maps are normal for this case study.

**Table 11** Simulation accuracies (percentages of correctly simulated pixels) of the optimal maps and individual classes using three different transiogram joint modeling methods for Case study two (the Scotland dataset)

| Transiogram Modeling method | Overall | Class 1 | Class 2 | Class 3 | Class 4 | Class 5 | Class 6 |
|---|---|---|---|---|---|---|---|
| Linear interpolation | 70.92 | 65.40 | 74.99 | 16.62 | 49.20 | 31.68 | 77.19 |
| Mathematical model | 70.76 | 65.20 | 74.27 | 15.13 | 49.72 | 30.87 | 78.52 |
| Mixed | 70.98 | 65.51 | 75.13 | 16.62 | 49.89 | 30.64 | 76.97 |

**Table 12** Proportions of individual classes in the reference map, random sample data, and three optimal maps for Case study two (the Scotland dataset)

|  | Class 1 | Class 2 | Class 3 | Class 4 | Class 5 | Class 6 |
|---|---|---|---|---|---|---|
| Reference map | 28.19 | 48.05 | 0.43 | 2.98 | 1.74 | 18.61 |
| Sample data | 29.31 | 47.04 | 0.45 | 2.77 | 1.44 | 18.99 |
| Linear interpolation | 26.92 | 46.06 | 0.25 | 2.54 | 1.44 | 22.79 |
| Mathematical model | 26.82 | 45.46 | 0.23 | 2.56 | 1.42 | 23.51 |
| Mixed | 26.97 | 46.18 | 0.25 | 2.63 | 1.39 | 22.58 |

## 5. Discussions

Comparing the results from the two case studies found that the three methods of transiogram joint modeling for MCRF simulation perform similarly in terms of overall accuracy and spatial pattern. Using mathematical models with expert knowledge for transiogram modeling of minor classes may ensure their representation in simulated realizations and potentially improve their simulation accuracies, when the sample data is relatively sparse and linear interpolation cannot generate reliable transiogram models for minor classes. However, when sample data is not much sparse and minor classes are not extremely small and can produce meaningful experimental transiograms, using linear interpolation for transiogram modeling can perform well.

There are several factors that lead to such results: (1) When there are enough sample data, the tolerance width of experimental transiograms can be set to a relatively small number. For example, in Case study one, the tolerance width of experimental transiograms for the dense samples was set to 3 pixel length, while for the sparse samples it was set to 7 pixel length. Therefore, using linear interpolation for transiogram modeling could allow capturing the complex variations of experimental transiograms. (2) Although there are some differences between transiogram models using linear interpolation and mathematical model fitting, both are based on experimental transiograms, which are directly calculated from sample data. (3) In fitting experimental transiograms with mathematical models, we emphasized the low lag section so that the low lag sections of experimental transiograms can be more respected by transiogram models, because only the low lag sections of transiogram models are needed in a sequential simulation. Towards the end of the simulation of a realization, only very close data points are used in calculating the local conditional probability distribution. Since the lower lag sections of transiogram models from different transiogram joint modeling methods are consistent for most transiogram models, it is not surprising that no significant accuracy differences were observed between simulated results using different transiogram joint modeling methods.

Despite the consistent performance in simulation accuracy, the used methods for acquiring transiogram models from experimental transiograms have their respective advantages and shortcomings. Using mathematical models to jointly fit experimental transiograms provides the condition to incorporate expert knowledge and it usually performs better when some experimental transiograms are not much reliable, but it is time-consuming. For example, it is not easy to properly determine the correlation range parameter and the model type if users do not have experience and the knowledge about the simulated landscape in the study area. Linearly interpolating experimental transiograms into continuous models is very efficient and able to capture subtle variations when sample data are sufficient, but it cannot incorporate expert knowledge.

This method is not reasonable to use when sample data are insufficient, especially for some extremely minor classes, because under such situation some experimental transiograms cannot be properly estimated. It can be seen that in Case study one (the Iowa dataset), transiogram models by mathematical model fitting and transiogram models by the mixed method can improve the simulation accuracy of the minor class 5, because the experimental transiograms of the minor class 5 are unreliable.

In a geostatistical simulation, simulated results, especially class proportions in simulated realizations, are decided by three factors – geostatistical model itself (here the MCRF model, for local conditional probability distribution estimation), sample data (for conditioning the simulation), and input parameters (here transiogram models, for representing spatial correlations). If the geostatistical model can fairly estimate all classes (both major and minor classes), then given a sample dataset for conditioning the simulation, it is the input parameters (transiogram models at here) that decide whether all classes can be fairly estimated in the simulated results. Therefore, estimating the transiogram models as objectively as possible based on sample data and expert knowledge is important.

It is quite normal that some classes may be relatively under/over-estimated/represented in simulation of categorical spatial variables to some extent due to various reasons in sampling, parameter evaluation and optimization. This should not be confused with the small class underestimation problem of the coupled Markov chain model. The latter is a proved theoretical defect caused by improper model construction (i.e., the independency assumption of the two perpendicular Markov chains causes conflict transitions which have to be excluded in computation), and it occurs in the estimation of local conditional probability distribution with fair parameter input; that is, given the same neighborhood and the same set of reasonable parameters, the coupled Markov chain model underestimates the occurrence probabilities of small classes (and consequently overestimates the occurrence probabilities of major classes) compared with other spatial models unless all classes have equal proportions (i.e., there is no small class) [11, 29]. Therefore, the small class underestimation of the coupled Markov chain model in simulated realizations (actually in local conditional probability distributions) is irrelevant with the occasional under/over-estimation/representation of some classes (major or minor classes) caused by other reasons (e.g., imperfections in sampling and parameter evaluation, or optimization), and it is far stronger due to the propagation effect in the simulation process.

It should be noted that the aim of this paper is to test the sensitivity of the MCRF model to the three different transiogram joint modeling methods rather than to make a highly accurate application study, thus all simulations were purely based on sample data and no ancillary data (e.g., legacy data, remote sensing data) were used for improving the simulation accuracies or simulated patterns. This is why our simulated maps only captured the general patterns of soil types but did not capture the details (especially the linear features of some soil patches) displayed in some areas of reference maps in both case studies. This insufficiency can be largely overcome in real applications when some ancillary datasets are incorporated. It is worth mentioning that using multiple-point statistics in spatial simulation can better reproduce the spatial patterns (e.g., curvilinear features), as shown by the multiple-point geostatistical approach [32], but training images are needed for estimating the multiple-point statistics and implementation is complex. The MCRF approach also can be a multiple-point statistical method because the full formula of the MCRF model contains multiple-point likelihood terms [18, 19]; this may be a future effort direction.

## 6. Conclusions

Transiogram joint modeling based on experimental transiograms estimated from sample data is essential to simulation of categorical spatial variables such as soil types and land cover classes, because it provides the MCRF model the transition probability parameters needed at any lag values that can meet the constraint conditions of transition probabilities. In this study, we detailed the three transiogram joint modeling methods based on experimental transiograms (i.e., linear interpolation, mathematical model joint fitting, and the mixed method) and tested their effects on simulated results of the MCRF model by simulating the spatial patterns of soil types using the MCSS algorithm. Our results indicate that when sample data are sufficient for estimating reliable experimental transiograms, the MCRF model is insensitive to the different transiogram model sets obtained using different joint modeling methods. Small differences in transiogram models do not affect the simulation accuracy significantly regardless of study area and area size. Although slight variations in realizations and optimal maps were observed between different transiogram joint modeling methods, the overall patterns and accuracies of simulated realizations and optimal maps were similar in the two case studies. However, when sample data are sparse and sample data of minor classes are too few to generate reliable experimental transiograms, the MCRF model can be sensitive to the transiogram models of minor classes in terms of the class accuracy. Under this situation, using mathematical models to tackle the unreliability problem of linear interpolation-based empirical transiogram models of minor classes is necessary to make sure the proper simulation of minor classes. Minor classes usually have lower simulation accuracy than major classes have because they tend to have larger uncertainty (i.e., occur more randomly). If a minor class is under-sampled compared to its due proportion (e.g., its proportion in a denser sample set or reference map), its simulation accuracy can be severely reduced, as shown by the class 5 in the sparse sample dataset of Case study one, because that largely increases the uncertainty of the simulated minor class.

This study also showed that the exponential model is a good choice to fit experimental auto-transiograms, and the exponential, Gaussian, spherical, and corresponding gamma distribution-based composite models are proper models to fit experimental cross-transiograms. When experimental transiograms are representative, linear interpolation can be an efficient way to obtain transiogram models.

However, when estimating reliable experimental transiograms is difficult from a very sparse sample dataset, one should use these mathematical models to obtain transiogram models, especially transiogram models of minor classes, by fitting them to experimental transiograms with expert judgment on transiogram model parameters.

**References**

1. Arnold, D.B., Milne, W.J.: The use of Voronoi tessellations in processing soil survey results. IEEE Computer Graphics and Applications, **4**(3), 22-28 (1984)
2. Bierkens, M.F.P., Burrough, P.A.: The indicator approach to categorical soil data. Journal of Soil Science, **44**(2), 369–381 (1993)
3. Carle, S.F., Fogg, G.E.: Transition probability-based indicator geostatistics. Mathematical Geology, **28**(4), 453–476 (1996)
4. Carle, S.F., Fogg, G.E.: Modeling spatial variability with one and multidimensional continuous-lag Markov chains. Mathematical Geology, **29**(7), 891–918 (1997)
5. Casella, G., Berger, R.L.: Statistical Inference, 2$^{nd}$ Edition. Duxbury Press, Pacific Grove (2002)
6. Deutsch, C.V., Journel, A.G.: GSLIB: Geostatistical Software Library and Users Guide. Oxford University Press, New York (1998)
7. Ethier, V.G.: Application of Markov analysis to the Banff formation (Mississippian), Alberta. Journal of the International Association for Mathematical Geology, **7**(1), 47–61 (1975)
8. Goovaerts, P.: Geostatistics in soil science: state-of-the-art and perspectives. Geoderma, **89**(1-2), 1–45 (1999)
9. Isaaks, E.H., Sarivastava, R.: An Introduction to Applied Geostatistics. New York, Oxford University Press (1989)
10. Jaffray, J.Y.: Bayesian updating and belief functions. IEEE Transactions on Systems, Man, and Cybernetics, **22**(5), 1144-1152 (1992)
11. Li, W.: Markov chain random fields for estimation of categorical variables. Mathematical Geology, **39**(3), 321–335 (2007a)
12. Li, W.: Transiograms for characterizing spatial variability of soil classes. Soil Science Society of America Journal, **71**(3), 881–893 (2007b)
13. Li, W., Zhang, C.: A random-path Markov chain algorithm for simulating categorical soil variables from random point samples. Soil Science Society of America Journal, **71**(3), 656–668 (2007)
14. Li, W., Zhang, C.: A single-chain-based multidimensional Markov chain model for subsurface characterization. Environmental and Ecological Statistics, **15**(2), 157-174 (2008)
15. Li, W., Zhang, C.: Linear interpolation and joint model fitting of experimental transiograms for Markov chain simulation of categorical spatial variables. International Journal of Geographical Information Science, **24**(6), 821–839 (2010)
16. Li, W., Zhang, C., Dey, D.K.: Modeling experimental cross-transiograms of neighboring landscape categories with the gamma distribution. International Journal of Geographical Information Science, **26**(4), 599–620 (2012)

17. Li, W., Zhang, C., Dey, D.K., Willig, M.R.: Updating categorical soil maps using limited survey data by Bayesian Markov chain cosimulation. The Scientific World Journal, **2013**, Article ID 587284 (2013)
18. Li, W., Zhang, C., Willig, M.R., Dey, D.K., Wang, G., You, L.: Bayesian Markov chain random field cosimulation for improving land cover classification accuracy. Mathematical Geosciences, **47**(2), 123-148 (2015)
19. Li, W., Zhang, C.: Markov chain random fields in the perspective of spatial Bayesian networks and optimal neighborhoods for simulation of categorical fields. Computational Geosciences, **23**(5), 1087–1106 (2019)
20. Luo, J.: Transition probability approach to statistical analysis of spatial qualitative variables in geology. In: Forster, A., Merriam, D.F. (eds.) Geologic Modeling and Mapping, pp. 281-299. Plenum Press, New York (1996)
21. Matheron, G.: Principles of geostatistics. Economic geology, **58**(8), 1246-1266 (1963)
22. Meerschman, E., Van Meirvenne, M., Van De Vijver, E., De Smedt, P., Islam, M.M., Saey, T.: Mapping complex soil patterns with multiple-point geostatistics. European Journal of Soil Science, **64**(2), 183-191 (2013)
23. Pearl, J.: Fusion, propagation, and structuring in belief networks. Artificial Intelligence, **29**(3), 241-288 (1986)
24. Schwarzacher, W.: The use of Markov chains in the study of sedimentary cycles. Mathematical Geology, **1**, 17-39 (1969)
25. Webster, R., Oliver, M.A.: Geostatistics for environmental scientists. John Wiley & Sons, Chichester (2007)
26. Yu, J., Li, W., Zhang, C.: A framework of experimental transiogram modelling for Markov chain geostatistical simulation of landscape categories. Computers, Environment and Urban Systems, **73**, 16-26 (2019)
27. Zhai, R., Li, W., Zhang, C., Zhang, W., Wang, W.: The transiogram as a graphic metric for characterizing the spatial patterns of landscapes. Landscape Ecology, **34**(9), 2103-2121 (2019)
28. Zhang, C., Li, W.: Regional-scale modelling of the spatial distribution of surface and subsurface textural classes in alluvial soils using Markov chain geostatistics. Soil use and management, **24**(3), 263-272 (2008a)
29. Zhang, C., Li, W.: A comparative study of nonlinear Markov chain models for conditional simulation of multinomial classes from regular samples. Stochastic Environmental Research and Risk Assessment, **22**(2), 217-230 (2008b)
30. Zhang, W., Li, W., Zhang, C.: Land cover post-classifications by Markov chain geostatistical cosimulation based on pre-classifications by different conventional classifiers. International Journal of Remote Sensing, **37**(4), 926-949 (2016)
31. Zhang, W., Li, W., Zhang, C., Zhao, T.: Parallel computing solutions for Markov chain spatial sequential simulation of categorical fields. International Journal of Digital Earth, **12**(5), 566-582 (2019)
32. Strebelle, S.: Conditional simulation of complex geological structures using multiple-point statistics. Mathematical Geology, **34**, 1-21 (2002)